\documentclass[a4paper,11pt]{article}
\usepackage{pos}
\usepackage{placeins}

\newcommand{\be}{\begin{equation}}
\newcommand{\ee}{\end{equation}}
\newcommand{\Det}{\mathrm{Det}}

\newcommand{\tr}{\mathrm{tr}}
\newcommand{\MF}{\mathrm{MF}}
\newcommand{\G}{\mathrm{G}}
\newcommand{\cl}{\mathrm{cl}}

\title{Naive Gaussian approximation in a quark-meson model}

\author*[a,c]{Győző Kovács}
\author[b]{Zsolt Szép}
\author[c]{Péter Kovács}
\author[c]{György Wolf}

\affiliation[a]{Institute of Theoretical Physics, University of Wroclaw, \\PL-50204 Wrocław, Poland}

\affiliation[b]{HUN-REN-ELTE Theoretical Physics Research Group,\\
Pázmány Péter sétány 1/A, 1117 Budapest, Hungary}

\affiliation[c]{Institute for Particle and Nuclear Physics, Wigner Research Centre for Physics,\\
  Konkoly-Thege Miklós út 29–33, 1121 Budapest, Hungary}

\emailAdd{kovacs.gyozo@wigner.hu}

\abstract{A sequence of approximations is derived to go beyond the mean-field level in the case of a simple linear sigma model. A naive, local version of the Gaussian approximation is discussed for the $2+1$ flavor extended Polyakov quark-meson model at finite temperature and chemical potential. Although at small chemical potential the pion mass has an unphysical thermal behavior in the usual parameterizations, it is shown that the pseudocritical temperature decreases in the presence of the thermal pion fluctuations. Compared to the mean-field level, the location of the CEP is only slightly modified, while the critical scaling around the CEP is unaffected.}

\FullConference{%
  FAIR next generation scientists - 8th Edition Workshop (FAIRness2024)\\
  23-27 September 2024\\
  Donji Seget, Croatia
}

\hypersetup{
pdftitle={Naive Gaussian approximation in a quark-meson model},
pdfsubject={Contribution to FAIRness 2024 proceedings},
pdfauthor={Gy. Kovacs, Zs. Szep, P. Kovacs, Gy. Wolf}
}

\begin{document}
\maketitle

\section{Introduction}

Effective field theoretical models are important tools for investigating the heavy-ion physics phenomenology. Approaches like the Nambo--Jona-Lasignio model, the quark-meson model, and their extended versions are frequently used to study the phase transition and critical phenomena in strongly interacting matter. Their usefulness stems from two features: First, they are applicable at finite temperature and chemical potential, where we lack first-principle calculations such as the lattice QCD. Second, they are easy to understand and implement. In most cases, these models are solved in a mean-field approximation, which provides a clear physical picture and a simple model setup, but sometimes suffer from the limitations of this approach.

In this work, we present a systematic way to go beyond the mean-field approximation in the case of a quark-meson model. The employed Gaussian approximation, which was already outlined in \cite{Kovacs:2021kas} and in the NJL context used in \cite{Yamazaki:2012ux}, includes the one-loop fluctuations of the fermion dressed meson fields in the form of a ring-resummation. We applied this approximation to an advanced, $N_f=2+1$ vector and axial vector meson extended Polyakov quark-meson model (ePQM) \cite{Kovacs:2016juc, Kovacs:2021kas}. In Ref.~\cite{Kovacs:2016juc} there was already an attempt to include the meson fluctuations, however, the corresponding contributions were only included in the pressure, after solving the field equations at mean-field level. Hence we present a consistent treatment of that approximation. 

\section{Quark-meson model at mean field and beyond}

In \cite{Kovacs:2021kas} a sequence of approximations was derived using the methods developed in \cite{Jackiw:1974cv}. Here we review these results to provide context for our calculations. We start with the simple $N_f=1$ model with a single scalar field, which is defined by the Lagrangian
\be
\mathcal{L}=\frac{1}{2}\partial_\mu\phi\partial^\mu\phi - V_\cl (\phi) + \bar\psi \left( i\gamma_\mu\partial^\mu -g \phi \right) \psi\, ,
\ee 
where $V_{\cl}(\phi)=\frac{1}{2} m^2 \phi^2 +\frac{\lambda}{4!}\phi^4$ is the classical potential.
The partition function can be written as
\begin{align} \label{eq:Z}
\mathcal{Z}=&\int \mathcal{D}\phi \mathcal{D}\bar\psi \mathcal{D}\psi \exp \left[ i\left( {S}_m (\phi) + {S}_f (\phi, \bar\psi, \psi) \right) \right] \nonumber \\
=& \int \mathcal{D}\phi \exp\left[ i \left( S_m (\phi) - i \log \Det \left(i\mathcal{S}^{-1} (\phi)\right) \right) \right]\, ,
\end{align}
where in the last step the integration over the fermionic fields was carried out in the usual way to obtain the fermionic functional determinant $\Det \left(i\mathcal{S}^{-1} (\phi)\right)$. At this point we are left with only the mesonic functional integral, which, however, cannot be evaluated analytically.

To obtain the different approximations one may use that the spontaneous symmetry breaking can be treated with the replacement $\phi(x)\to\bar\phi+\varphi(x)$, where $\bar\phi\equiv\langle\phi\rangle$ is a homogeneous background, while $\varphi(x)$ is the fluctuating mesonic field. In the simplest scenario, the fluctuating fields are set to zero. Then, the functional integral is trivial and one obtains the mean-field effective potential
\be 
V_{\MF}(\bar \phi)= V_{\cl}(\bar \phi) + i \int_k \log \det \left(i\mathcal{S}_0^{-1} (k)\right)\, ,
\ee 
where $i\mathcal{S}_0^{-1}$ is the inverse fermion propagator with vanishing fluctuating meson field.
To add corrections to the classical result, following \cite{Jackiw:1974cv, Kovacs:2016juc} the full action $S(\phi)$ in the second line of Eq.~\eqref{eq:Z} (the quantity in the round brackets) is expanded in the meson fields. The first correction comes from the quadratic term
\be \label{eq:S_expansion}
S_\mathrm{corr}(\bar\phi,\varphi) = \int_x \int_y \varphi(x) ~\frac{\delta^2 S(\phi)}{\delta \phi(x) \delta \phi(y)} \Big|_{\phi=\bar\phi}~\varphi(y)\, .
\ee 
The \textit{ideal-gas} approximation is obtained by retaining only $S_m(\phi)$ in Eq.~\eqref{eq:S_expansion}.
However, by using the full action, we arrive at what might be called the \textit{Gaussian} approximation. In both cases, the remaining Gaussian functional integral for the meson fields can be evaluated to obtain
\be 
V_\G = V_\MF - \frac{i}{2} \int_k \log \det (i{G}^{-1}(k))\, .
\ee 
Here $i{G}^{-1}(k)= i\mathcal{D}^{-1}(k)$ is the tree-level inverse meson propagator for the ideal-gas approximation, while $i{G}^{-1}(k)=i\mathcal{G}^{-1}(k)\equiv i\mathcal{D}^{-1}(k)-\Pi(k)$ for the Gaussian approximation, with $\Pi(k)$ being the fermionic one-loop contribution to the meson self-energy. In the latter case, it can also be shown by expanding the logarithm that $V_\G$ includes a ring resummation, with $\Pi$ petals connected by the tree-level meson propagators.

In this work, we use the Gaussian approximation with a local meson self-energy, in which case 
\be 
i\mathcal{G}^{-1}(k)=i\mathcal{D}^{-1}(k)-\Pi(0)=k^2-M^2\, ,
\ee 
where $M^2=\partial^2V_\MF/\partial\varphi^2\big|_{\phi=\bar\phi}$ is the mean-field curvature meson mass that includes the fermion one-loop correction. However, by calculating the curvature mass in the Gaussian approximation from $V_\G$ and using it to parameterize the model, the meson mass in the mesonic correction to the pressure would be unphysical. This highlights the challenge of this approach, as curvature masses
of various order appear in the calculation. A possible solution would be to relate the meson masses (or other quantities) at subsequent levels of the approximation and to impose constraints on the parameter space of the model by these relations. For the present paper, we trivially use the mean-field curvature masses for the parameterization and apply the Gaussian approximation to compute the pressure and the field equation. 

\section{Naive application of the local Gaussian approximation}

We turn to the implementation of the local Gaussain approximation in the ePQM model. The grand potential might be written in the form
\be \label{eq:Omega_4D}
\Omega_\G = \Omega_\text{tree}+i \tr\int_k \log (i\mathcal{S}_0^{-1} (k)) -i\frac{1}{2} \int_k \log \det (i\mathcal{G}^{-1}(k) )\, ,
\ee 
where the first two terms -- the tree-level and the fermionic one-loop contribution -- give the mean-field level grand potential $\Omega_\MF$.
As discussed concerning the (pseudo)scalar-(axial-)vector mixing in Ref.~\cite{Kovacs:2021kas} the mesonic functional determinant factorizes into a product of the inverse propagators of the physical modes. Therefore, using $\tr\log=\log\det$, the last term of \eqref{eq:Omega_4D} can be written as a sum of the contribution from the different modes. With this observation, after carrying out the Matsubara sum one obtains
\be \label{eq:Omega_cor}
\Omega_\G = \Omega_\MF + \sum_b n_b \int \frac{d^3k}{(2\pi)^3} \left( \frac{E_{(b)k}}{2}+ T \log \left( 1-e^{-\beta E_{(b)k} }\right)\right)\, ,
\ee
where $n_b$ is the multiplicity factor of a given physical mode. In the isospin symmetric limit one has $n_b=3$ for the isotriplet ($1-3$ flavor sector), $n_b=4$ for the isodoublets ($4-7$ flavor sector) together, and $n_b=1$ for both of the two physical isosinglet states ($N-S$ flavor sector) in each nonet. From Eq.~\eqref{eq:Omega_cor} one can get the field equations for the chiral condensates $\phi_N$ and $\phi_S$\footnote{The field equations for the Polyakov loop variables remain the same as at the mean-field level, hence we do not discuss them separately.}
\be \label{eq:FE}
0=\frac{\partial \Omega_G}{\partial \phi_{N/S}}  = \frac{\partial \Omega_\MF}{\partial \phi_{N/S}} + \sum_b  n_b  \frac{\partial M_b^2}{\partial \phi_{N/S}} \int \frac{d^3k}{(2\pi)^3} \frac{1}{2 E_{(b)k}}\left( \frac{1}{2}+ \frac{1}{e^{\beta E_{(b)k} }-1} \right)\, ,
\ee
where $M_b^2$ is the mean-field curvature mass of the meson $b$. These masses are listed in Ref.~\cite{Kovacs:2016juc} and, more generally, also in Ref.~\cite{Kovacs:2021kas}. It should be noted that the divergent mesonic vacuum fluctuation needs regularization, which can easily be done by, e.g., using dimensional regularization or a simple cutoff. The proper renormalization of this contribution may be more challenging, but we do not discuss this problem here.

For simplicity, in the present work only the pion fluctuations are considered. One might extend this scenario by including also the sigma meson and the kaons, but the contributions of heavier mesons are expected to be very small, if not negligible. Furthermore, we completely omit the mesonic vacuum contribution. By doing so, we aim to make a more consistent implementation of the approximation used in \cite{Kovacs:2016juc} for the pressure by applying the Gaussian approximation also in the field equations. Moreover, in this case the mean-field and the Gaussian vacuum meson masses are equal, and hence are equivalent for the parameterization of the model. 

\begin{figure}[htb]
    \centering
    \includegraphics[width=0.48\linewidth]{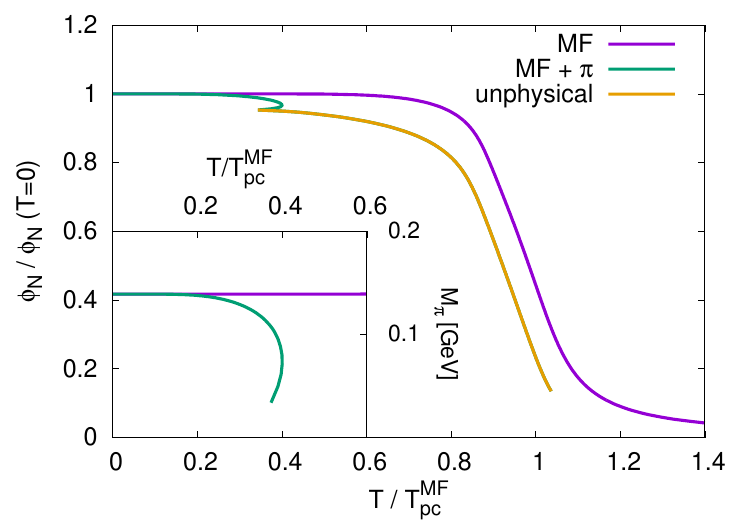}
    \includegraphics[width=0.48\linewidth]{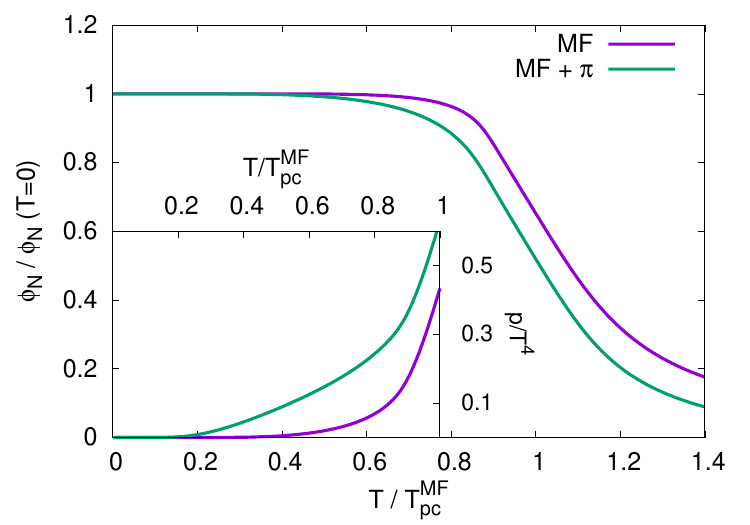}
    \caption{The temperature dependence of the nonstrange meson condensate $\phi_N$ in the mean-field and in the Gaussian approximation using parameter sets with physical (left panel) and increased (right panel) vacuum pion mass. The inset on the left panel shows how $M_\pi$ breaks down, while the inset on the right panel shows the enhancement of the pressure due to the meson fluctuations below $T_{pc}$.}
    \label{fig:order-param}
\end{figure}
At finite temperatures, the mesonic and the fermionic thermal fluctuations are expected to have a similar effect in driving the chiral symmetry restoration.\footnote{For the meson fluctuations this depends also on the sign of $\partial M_b^2/\partial \phi_{N/S}$, which is negative in some specific cases.}
However, since the pion mass is much smaller than the constituent quark masses, the thermodynamics might suffer from the early onset of the meson fluctuations. If the pion is light, the decrease of the chiral condensates $\phi_N$ and $\phi_S$ at low temperatures leads to a strong decrease in $M_\pi^2$. Thus, the pion becomes even lighter, which amplifies this process until $M_\pi^2$ drops to zero. This can be seen at $\mu_q=0$ in the left panel of Fig.~\ref{fig:order-param}, where we used the parameter set presented in Ref.~\cite{Kovacs:2021kas}, but similar behavior can be found with the parameterizations of Refs.~\cite{Kovacs:2016juc} and \cite{Kovacs:2021ger}. In the unphysical case when $M_\pi^2$ is negative $\left|M_\pi^2\right|$ is used. The unphysical behavior of the pion mass is depicted in the inset of Fig.~\ref{fig:order-param}. The complete solution to this problem is not trivial, but is generally possible. However, it is beyond the scope of this paper and will be discussed elsewhere. For the present work, we simply use an alternative parameterization obtained from the set in Ref.~\cite{Kovacs:2021kas} by increasing the value of the bare mass parameter for the (pseudo)scalar fields to $m_0^2=0.03$~GeV$^2$, which increases the vacuum mass of the pion to $M_{\pi,\mathrm{mod}}\approx 2.5 M_{\pi,\mathrm{phys}}$. 
Although the masses are shifted from their physical value, the effect of the mesonic thermal fluctuation can be investigated in this scenario, which is depicted in the right panel of Fig.~\ref{fig:order-param}. As expected, the transition temperature decreases due to the presence of the meson fluctuations. Moreover, as shown in the inset of Fig.~\ref{fig:order-param}, the contribution of the pions -- which we lack in the mean-field approximation -- increases the pressure similarly as in Ref.~\cite{Kovacs:2016juc}.

\begin{figure}[htb]
    \centering
    \includegraphics[width=0.48\linewidth]{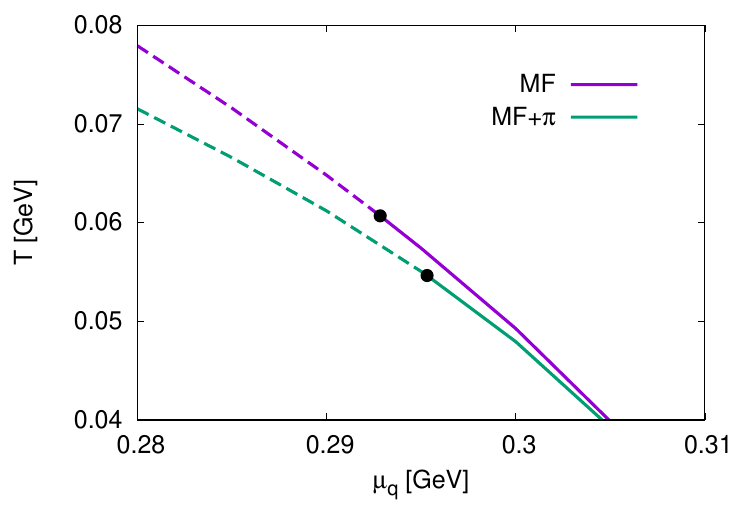}
    \includegraphics[width=0.48\linewidth]{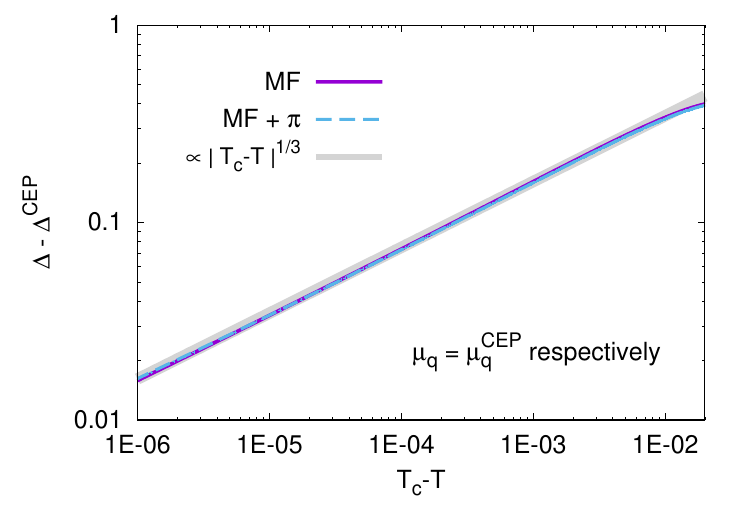}
    \caption{The critical endpoint in the mean-field and the Gaussian approximation with the parameter set of Ref.~\cite{Kovacs:2021kas} (left) and the critical scaling for the subtracted condensate $\Delta$ (right). 
    }
    \label{fig:CEP}
\end{figure}
At large chemical potentials, where the CEP might be found, $M_\pi^2$ remains positive. This is expected, since -- in contrast to the meson fluctuations -- the fermion thermal fluctuations explicitly depend on $\mu_q$ and thus can compensate for the decrease of the pion mass even at low temperatures if $\mu_q$ is sufficiently large. Therefore, we can study the CEP and the critical behavior in the Gaussian approximation also with a physical parameterization. As shown in the left panel of Fig.~\ref{fig:CEP}, the CEP shifts only slightly to lower temperatures and higher chemical potentials due to the effect of mesonic one-loop correction. For model parameters such that the original location of the CEP is at higher $T$ and lower $\mu_q$, this shift is also larger. 

To discuss the critical scaling, we define the so-called subtracted condensate as $\Delta=(\phi_N-\frac{h_N}{h_S}\phi_S)/(\phi_N^0-\frac{h_N}{h_S}\phi_S^0)$, which is used as an order parameter for the chiral symmetry breaking. To see the scaling of the order parameter with the critical exponent $\beta$, one has to approach the CEP rather precisely parallel to the phase boundary \cite{Kovacs:2006ym}. Consequently, if we approach at a fixed $\mu_q=\mu_q^\mathrm{CEP}$ in $T$, we expect to find the relation $\Delta-\Delta^\mathrm{CEP}\propto( T-T^\mathrm{CEP})^{1/\delta}$ corresponding to the direction orthogonal to the phase boundary. As shown in Fig.~\ref{fig:CEP} we indeed find such a scaling relation with $\delta=3$. This corresponds to the mean-field universality class, which therefore does not change in the local Gaussian approximation. 

\section{Conclusion}

We discussed a naive, local Gaussian approximation in the framework of a quark-meson model. We found that the mesonic thermal fluctuations drive the system towards the chiral restoration. However, at small chemical potential their effect is too strong and causes an unphysical behavior as $M_\pi^2$ becomes negative. At large chemical potentials these effects are absent. The presence of the meson fluctuations shifts the CEP to lower $T$ and higher $\mu_q$ compared to the mean-field approximation, but does not change the critical behavior of the model.

The presented approximation can, and should be improved in multiple directions. First, by including the mesonic vacuum fluctuations that might also partially compensate the effect of the mesonic thermal fluctuations. Second, the relation between the curvature masses defined in the mean-field and the Gaussian approximation could be investigated. These relations may be used also to constrain the model and to correct the irregular behavior of the pion mass. Finally, relaxing the local approximation could lead to interesting results by including nonlocal interactions in the meson fluctuations.

\vspace{.1cm}
\section{Acknowledgement}

This research was supported by the Hungarian National Research, Development and Innovation Fund under Project number K 138277. The work of Gy.~K. is partially supported by the Polish National Science Centre (NCN) under OPUS Grant
No. 2022/45/B/ST2/01527.

\FloatBarrier

\bibliographystyle{JHEP}
\bibliography{bib}

\end{document}